\documentclass[12pt,preprint]{aastex}
\usepackage{graphicx}
\usepackage{amssymb}
\usepackage{natbib}
\bibpunct{(}{)}{;}{a}{}{,}

\newcommand{\kms}  {km~s$^{-1}$}

\newcommand{\NGC}{NGC~7538}

\slugcomment{2008 October 29}

\shorttitle{12~GHz Maser Parallax of Cep~A and \NGC}
\shortauthors{Moscadelli et al.}

\begin{document}

\title{Trigonometric Parallaxes of Massive Star Forming Regions: II. \\
Cep~A \& \NGC}


\author{L. Moscadelli\altaffilmark{1}, M. J. Reid\altaffilmark{2}, 
        K. M. Menten\altaffilmark{3}, A. Brunthaler\altaffilmark{3}, 
        X. W. Zheng\altaffilmark{4} and Y. Xu\altaffilmark{3,5}}

\altaffiltext{1}{INAF, Osservatorio Astrofisico di Arcetri, 
   Largo E. Fermi 5, 50125 Firenze, Italy}
\altaffiltext{2}{Harvard-Smithsonian Center for
   Astrophysics, 60 Garden Street, Cambridge, MA 02138, USA}
\altaffiltext{3}{Max-Planck-Institut f\"ur Radioastronomie,
   Auf dem H\"ugel 69, 53121 Bonn, Germany}
\altaffiltext{4}{Department of Astronomy, Nanjing University
   Nanjing 210093, China}
\altaffiltext{5}{Purple Mountain Observatory, Chinese Academy of
   Sciences, Nanjing 210008, China}

\begin{abstract}
We report trigonometric parallaxes for the sources
\NGC\ and Cep~A, corresponding to distances of $2.65^{+0.12}_{-0.11}$~kpc 
and $0.70^{+0.04}_{-0.04}$~kpc, respectively.
The distance to \NGC\ is considerably smaller than its kinematic distance
and places it in the Perseus spiral arm.  
The distance to Cep~A is also smaller than its kinematic
distance and places it in the ``Local'' arm or spur.
Combining the distance and proper motions with observed
radial velocities gives the location and full space motion
of the star forming regions.
We find significant deviations from circular Galactic orbits
for these sources: both sources show large peculiar motions
($>$ 10~\kms) counter to Galactic rotation and \NGC\ has
a comparable peculiar motion toward the
Galactic center.  
\end{abstract}


\section{Introduction}
This paper is the second in a series of papers that
describe the results of a large program to determine Galactic structure
by measuring trigonometric parallaxes and proper motions.
Trigonometric parallaxes provide the ``gold standard'' of distance
measurements and can resolve fundamental questions of source
luminosity, mass, and age. 
 
The targets of our program are methanol (CH$_3$OH) masers associated
with high-mass star forming regions.  We used the National Radio 
Astronomy Observatory's 
\footnote{The National Radio Astronomy Observatory is a facility of the
National Science Foundation operated under cooperative agreement by
Associated Universities, Inc.}
Very Long Baseline Array (VLBA) to conduct
astrometic observations of the masers relative to compact extragalactic
radio sources, and we have achieved parallax accuracies approaching
$\pm$10~$\mu$as.   Background information about our program is given
in \citet{Rei08}, hereafter called Paper I.

In this paper we report VLBA observations of 12 GHz methanol masers
toward \NGC\ and Cep~A.  These well-studied sources are in the 
$2^{nd}$ quadrant of the Galaxy.  \NGC\ has a kinematic distance
of about 5.6~kpc, which would place it well past the Perseus
spiral arm, possibly in the ``Outer'' (``Cygnus'') arm.  However,
\citet{Xu06} found that the kinematic distance for W3OH, 
a source at a comparable Galactic longitude and kinematic 
distance, was a factor of two too great.  So, obtaining a direct
distance estimate is important to locate this source in the Galaxy.
While Cep~A is likely in the ``Local'' arm or spur, its distance is 
uncertain, with estimates ranging between 0.3~kpc \citep{Mig92} and 0.9~kpc \citep{Mor93}.  
Here we present parallax measurements of \NGC\ and Cep~A.

\section{Observations and Data Reduction}

Paper I describes the general observational setup and method of calibration.
Here we give only procedures and parameters specific to the observations of 
\NGC\ and Cep~A.  We used the VLBA (program BR100C) to observe the 
$2_0-3_{-1}$ E ($\nu_0 = 12178.597 $~MHz) transition of methanol
at five dates: 2005 September 9 and December 1 and 2006 February 25, May 26 
and September 1.  These dates were selected to symmetrically sample both
the eastward and northward parallax signatures and minimize correlations 
among the parallax and proper motion parameters.  The first, fourth and fifth 
observations were fully successful with a very low ($<$1\%) level of observing 
downtime; for the second epoch the Brewster antenna did not produce
fringes and for the third epoch Hancock did not observe due to bad weather.

In order to provide independent measures of parallax and reduce the risk of
structural variability of the background source, we used two background
continuum sources: J2254+6209 and J2302+6405.  Table~\ref{obs_sou} lists the positions of the masers 
and the background sources.
The dual circularly polarized 4~MHz bands containing the maser signals were
centered at LSR velocities ($V_{\rm LSR}$) of $-10$~\kms\ and $-60$~\kms\ for 
Cep~A and \NGC, respectively.  Spectral resolution was 0.38~\kms.

\begin{deluxetable}{lllcrcc}
\tablecolumns{6} \tablewidth{0pc} \tablecaption{Positions and Brightnesses}
\tablehead {
  \colhead{Source} & \colhead{R.A. (J2000)} &  \colhead{Dec. (J2000)} &
  \colhead{$\phi$}  & \colhead{P.A.} &
  \colhead{Brightness} & \colhead{V$_{\rm LSR}$} 
\\
  \colhead{}       & \colhead{(h~~m~~s)} &  \colhead{(\degr~~'~~'')} &
  \colhead{(\degr)}  & \colhead{(\degr)} &
  \colhead{(Jy/beam)} & \colhead{(\kms)} 
            }
\startdata
\NGC\      &  23 13 45.3622 & 61 28 10.507 & &  & 5--6 & $-$57  \\
J2254+6209  &  22 54 25.2930 & 62 09 38.725 & 2.4 & $-$73 & 0.07 &  \\
J2302+6405 &   23 02 41.3150 & 64 05 52.849  & 2.9 & $-$27 & 0.12 &  \\
                    &                            &                      &  &  &  &      \\
Cep~A &   22 56 18.0970 & 62 01 49.399 & & & 0.6--1 & $-$10 \\
J2254+6209  &  22 54 25.2930 & 62 09 38.725 & 0.3 & $-$59 & 0.07 &  \\
J2302+6405 &   23 02 41.3150 & 64 05 52.849  & 2.2 & 20 & 0.12 &  \\ 
\enddata
\tablecomments  {$\phi$ and P.A. are the separations and position angles (East of North) from the maser to the reference sources.
For both maser and quasar images, we used a circular restoring beam of 2~mas FWHM.
For both maser targets, the derived absolute position of the reference maser channel is based on the position of the quasar J2302+6405
 from the VLBA calibrator survey. The error in the absolute position of \NGC\ is dominated by the uncertainty in the absolute position of the quasar J2302+6405,
0.26~mas and 0.48~mas in R.A. and Dec., respectively.  For the weaker and extended Cep~A maser (see discussion in Sect.~\ref{fit_Cep}), 
the R.A. position error is $\sim$1.3~mas.}
\label{obs_sou}
\end{deluxetable}

We used observations of the strong VLBA calibrator 3C~454.3 to correct 
for instrumental delays and phase offsets among different frequency bands.
The spectral channel with the strongest maser emission was used as the 
phase reference: $V_{\rm LSR} = -4.2$~\kms\ for Cep~A and $V_{\rm LSR} = -55.8$~\kms\ 
for \NGC.  These reference features were detected at all epochs and were 
relatively stable, varying in intensity by less than $\pm20$\%.

For the maser data, after phase referencing we 
produced naturally-weighted maps for each spectral channel, covering
a region of $\approx$~2$\arcsec$.  Our spectral resolution was inadequate to resolve some narrow 
maser features and, to minimize spectral sidelobes, 
we Hanning smoothed the visibilities, reducing the velocity resolution 
to $\approx$~0.8~\kms.  
We searched all maser spectral channels 
for emission above a conservative threshold taken as the absolute value 
of the minimum in the map.  This allows for dynamic range limitations, 
as opposed to using a strict map-noise limit.  
%Our detections 
%were always greater than $7\sigma$, where $\sigma$ is the rms map noise 
%(evaluated in a region with no emission).  Values of $\sigma$ varied among 
%channels and epochs between 2 and 15~mJy~beam$^{-1}$, except for Cep~A 
%at the second epoch where it varied between 10 and 40~mJy~beam$^{-1}$. 
The detected maser spots were fitted with an elliptical 
Gaussians brightness distribution (using the AIPS task JMFIT). 

The maser images for both sources were 
elongated and extended over a size of a few mas.  This precluded using 
VLBA antennas that produced only long baselines that fully resolved the 
masers, since reference-phase solutions could not be obtained. Thus, 
compared to some other sources in our program, these observations 
had lower angular resolution.  For both maser sources we used 
a circular restoring beam of 2~mas FWHM, which was close to the 
interferometer ``dirty" beam.

The reference channel image for \NGC\ revealed some asymmetric structure 
(see Fig.~\ref{NGC_images_1}).  Thus, we self-calibrated (amplitude and phase)
the reference channel visibilities and applied the corrections to 
both the continuum and line data before mapping.  This preserves the
astrometric precision for relative position measurements.
The Cep~A reference channel was not strong enough to allow self-calibration.  
The image (see Fig.~\ref{Cep_images_1}) shows symmetric low-level structures
that are probably caused by small amplitude calibration errors.  Since,
for brightnesses $>20$\% of the peak, the source structure is 
relatively simple, resonable position accuracy could still be achieved.

For the background continuum sources (J2254+6209 and J2302+6405), 
we integrated the data from all four dual-polarized bands and imaged the 
sources using the AIPS task IMAGR.  
The naturally-weighted "dirty" beam was determined by the availability
of maser phase-reference data and was almost circular with a FWHM of
$1.9 \times 1.8$~mas.  Matching the maser images, we adopted a circular 
restoring beam of 2~mas (FWHM).

\section{Parallaxes and Proper Motions}

We measured the parallax and proper motions of the 12~GHz masers
from the change in the position differences of the masers with respect to the 
background continuum sources.  The change in position of a maser spot 
relative to the background source was modeled as a combination of
the parallax sinuisoid and a secular proper motion in each celestial 
coordinate.   See Paper~I for details of this procedure.

Since systematic errors, owing to maser blending, potential structure in
the background continuum sources, and unmodeled atmospheric delay variations
usually dominate over signal-to-noise limitations, we adopted an 
empirical approach to the weighting of the data.  We added ``error floors''
in quadrature with the formal position uncertainties, separately 
to the east and north position offsets.  These error floors were adjusted
until the residuals of parallax and proper motion fit yielded a
$\chi^{2}$ per degree of freedom near unity in each coordinate.

\subsection{\NGC}    \label{fit_NGC}

For \NGC, we found that spatial blending of maser features limited the 
accuracy of spot positions.  We attempted to fit multiple spatial components 
to blended images, but this yielded component positions with 
poor accuracies ($\ge0.1$~mas).  Instead we found that using the
the centroid position of the maser reference channel, which by definition is 
zero after phase referencing, improved the parallax fits compared to
multi-component fits.

\begin{figure}
\centering
\includegraphics[angle=-90.0,scale=0.8]{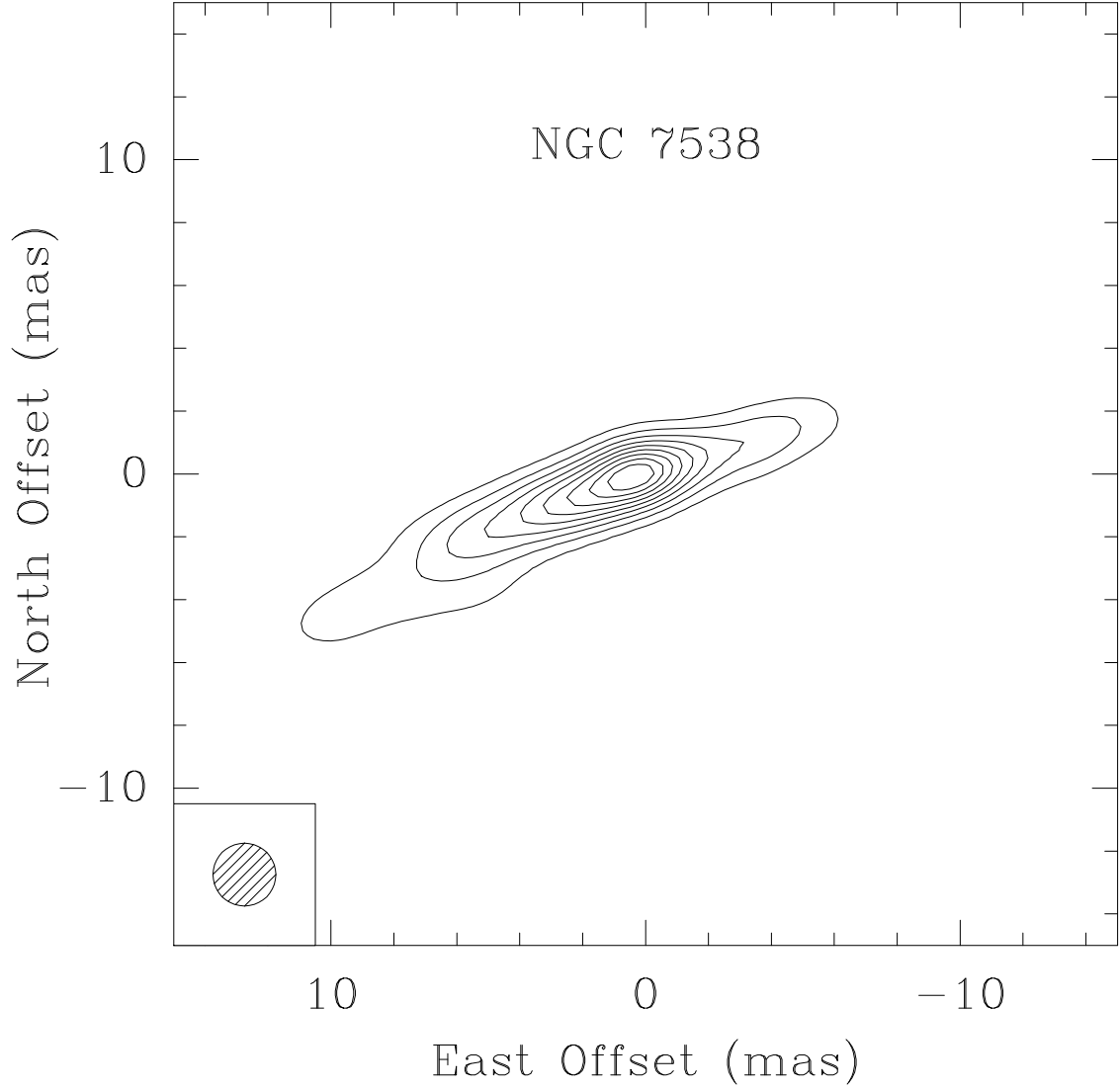}
\caption{Self-calibrated image of the reference maser channel 
at $V_{\rm LSR} = -55.8$~\kms\ of \NGC\ 
 for the first epoch (2005 Sep. 9). 
Contour levels are at multiples of 10\% of the peak brightness
of 6.2~Jy~beam$^{-1}$. 
The insert in the lower left shows the restoring beam.
\label{NGC_images_1}}
\end{figure}

\begin{figure}
\includegraphics[angle=0.0,scale=1.0]{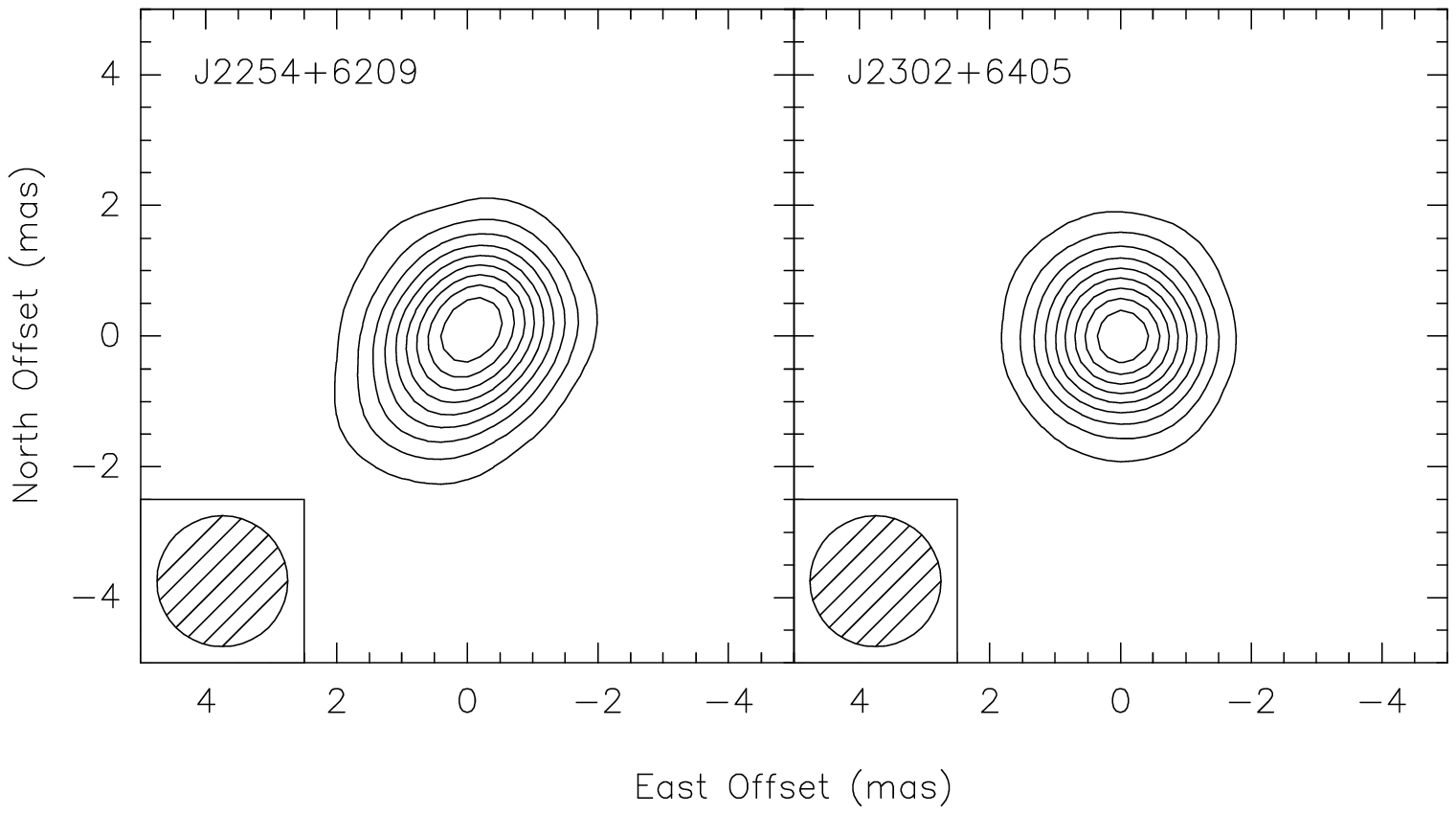}
\caption{Images of the two background continuum sources phase referenced to the \NGC\ maser. 
Source names are in the upper left corner and restoring beams are in the lower
left corner of each panel. Both images are from 
the first epoch observations on 2005 Sep. 9.  
Contour levels of both images are at multiples of 10\% of the peak brightness
of 0.07~Jy~beam$^{-1}$ and 0.12~Jy~beam$^{-1}$ 
for the J2254+6209 and the J2302+6405 image, respectively. 
\label{NGC_images_2}}
\end{figure}

The reference channel emission for \NGC\ (see Fig.~\ref{NGC_images_1}) had a peak brightness of 
$\approx6$~Jy~beam$^{-1}$, which was strong enough to produce good 
quality reference-phase solutions and yielded reasonable images
of the background sources J2254+6209 and J2302+6405 
(see Fig.~\ref{NGC_images_2}).  
Formal fitting uncertainties for the positions of the background sources
were $\sim0.01$~mas.

\begin{figure}
\includegraphics[angle=-90,scale=0.7]{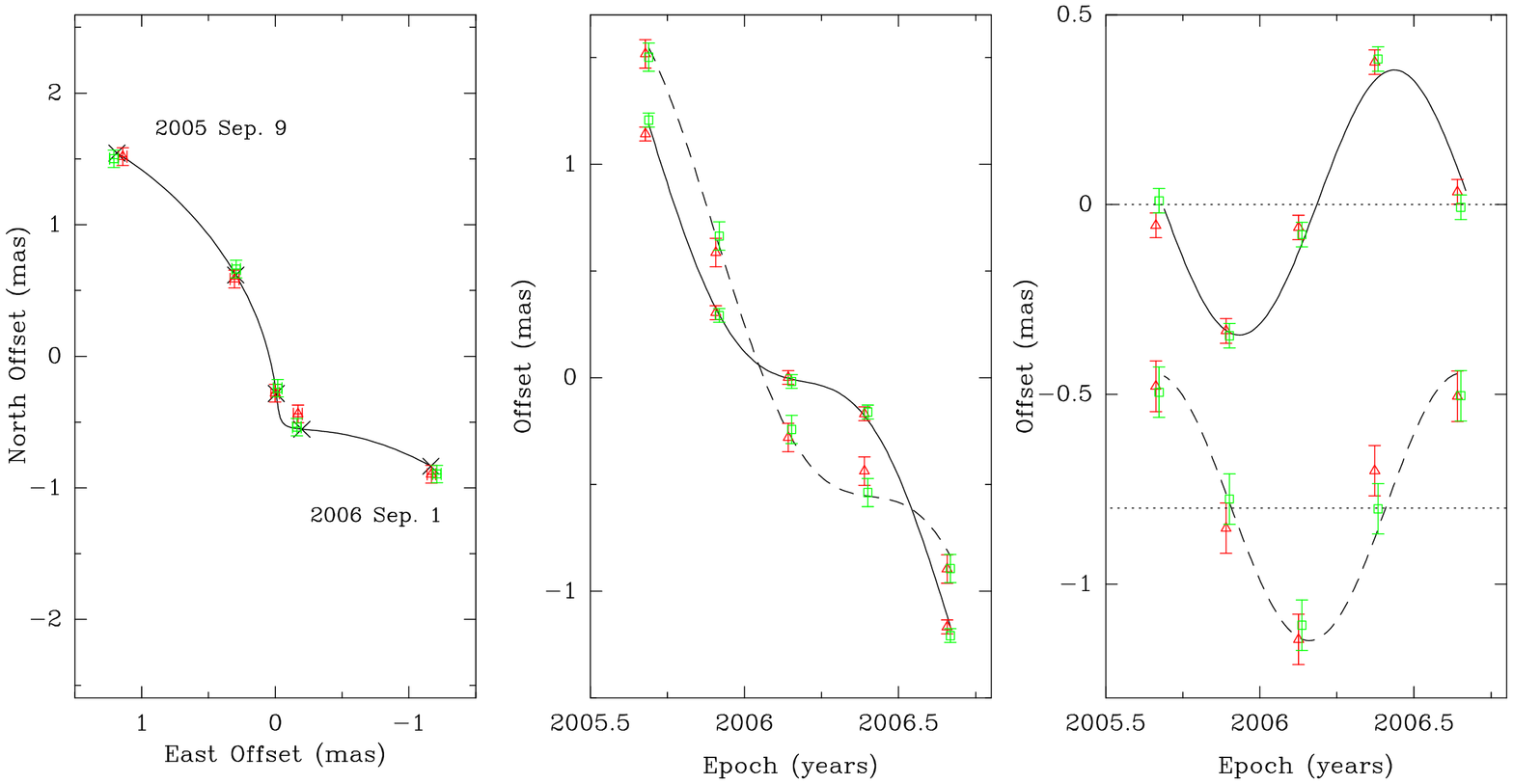}
\caption{Results of the parallax fit for \NGC.  For each plot, 
colored symbols indicate the position offset of the
reference maser channel centroid; the errorbars have been
scaled to give a reduced $\chi^2$ of unity, as discussed in the text. 
{\it Red triangles} and {\it green squares} refer to measurements
relative to the background quasars J2254+6209 and J2302+6405, respectively.  
{\bf Left Panel:}  Sky-projected motion of the maser. 
The {\it crosses} and the {\it continuous line} show the best-fit position
offsets and the trajectory, respectively.  The observing date of the 
first and last position are indicated. 
{\bf Middle Panel:} The position offsets of the maser along the East  
and North direction versus time.  The best-fit model of the variation of the 
East and North offsets with time is shown as
{\it continuous} and {\it dashed} lines, respectively.  
{\bf Right Panel:}  Same as for the middle panel, but with the fitted slopes 
(proper motions) subtracted. The {\it dotted} lines indicate zero position offset.
To avoid overlapping, the North offset data and model have been shifted to negative offsets.
\label{NGC7538_parallax}}
\end{figure}

\begin{deluxetable}{ccrrr}
\tablecolumns{6} \tablewidth{0pc} \tablecaption{\NGC\ : Parallax \& Proper Motion Fit}
\tablehead {
  \colhead{Maser V$_{\rm LSR}$} & \colhead{Background} &  \colhead{Parallax} &
  \colhead{$\mu_{x}$} & \colhead{$\mu_{y}$} 
\\
  \colhead{(km~s$^{-1}$)}  &  \colhead{Source} &  \colhead{(mas)} &
  \colhead{(mas~y$^{-1}$)} & \colhead{(mas~y$^{-1}$)} 
            }
\startdata
$-$55.8 & J2254+6209 & 0.371$\pm$0.026 & $-$2.41$\pm$0.05 & $-$2.41$\pm$0.12  \\
$-$55.8 & J2302+6405 &  0.414$\pm$0.019 &  $-$2.52$\pm$0.03 &  $-$2.47$\pm$0.13 \\
& & & & \\
$-$55.8 & \mbox{combined} &  0.378$\pm$0.017 &  $-$2.45$\pm$0.03 &  $-$2.44$\pm$0.06 \\
\enddata
\tablecomments {Col.~1 reports the LSR velocity of the reference maser channel; 
Col.~2 indicates the background quasar whose data were used for the parallax fit: "combined" means that both quasars' data were used; 
Col.~3 reports the fitted parallax; Cols.~4~and~5 give the fitted proper motions along the East and North direction, respectively.}
\label{par_fit_N538}
\end{deluxetable}

For \NGC, we first perfomed the parallax fit separately for the two 
background sources, J2254+6209 and J2302+6405. 
Next we produced a combined solution,
which required fewer total parameters, since we constrained the solutions
to have the same proper motion for the maser with respect to both background 
sources (as the background sources should have essentially zero proper motion).
The error floors, which account for systematic errors
in the relative postions, were  0.033~mas and 0.067~mas  for the eastward 
and northward directions, respectively.
Table~\ref{par_fit_N538} reports the results of these fits and
Fig.~\ref{NGC7538_parallax} shows the data and the best fitting models. 
The errorbars for the positions in this figure include the error floors.

\subsection {Cep~A}  \label{fit_Cep}

\begin{figure}
\centering
\includegraphics[angle=-90.0,scale=0.8]{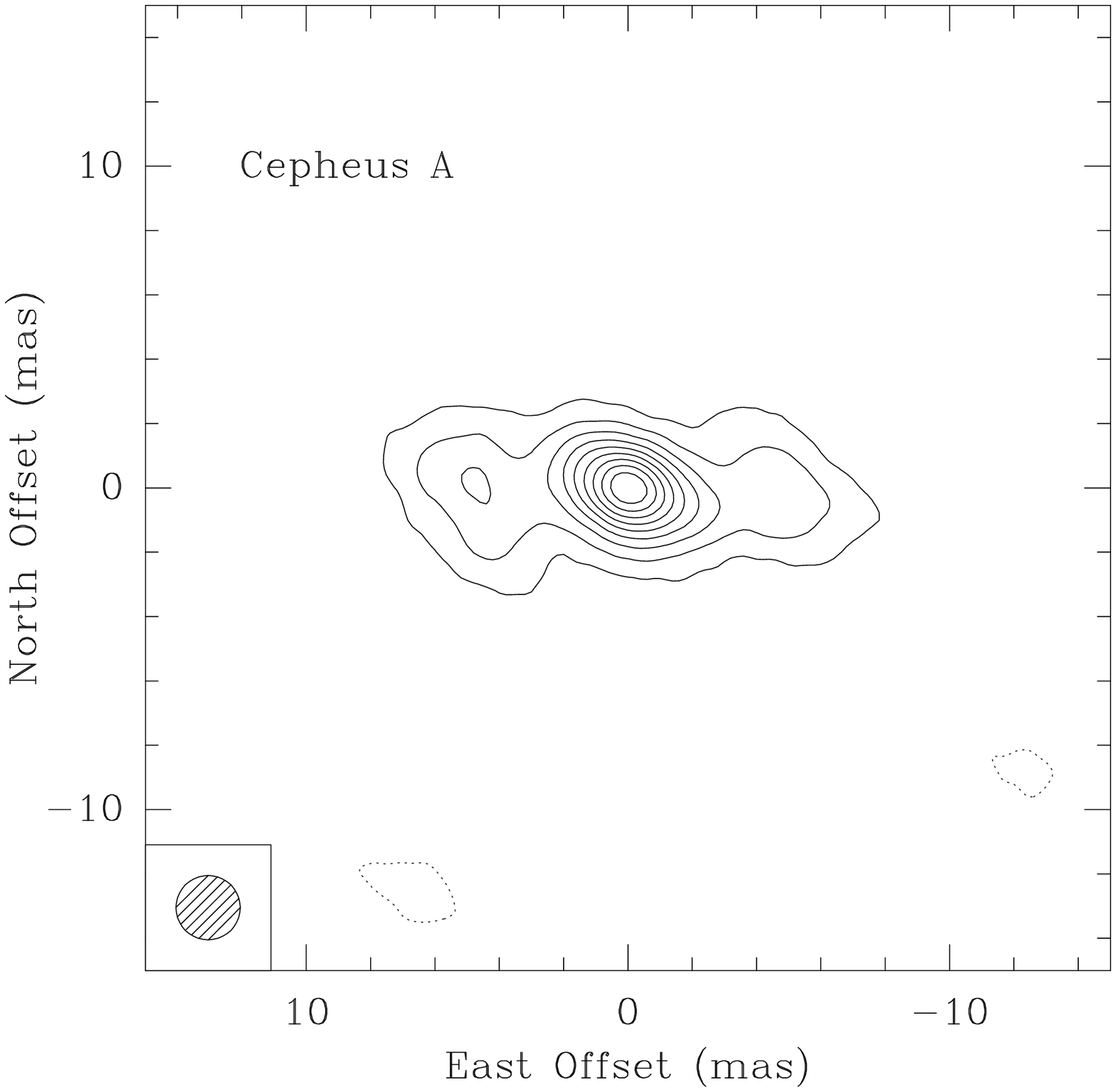}
\caption{Image of the reference maser channel at $V_{\rm LSR} = -4.2$~\kms\ 
of Cep~A for the first epoch (2005 Sep. 9). 
Contour levels are at multiples of 10\% of the peak brightness
of 0.9~Jy~beam$^{-1}$. 
The insert in the lower left shows the restoring beam.
\label{Cep_images_1}}
\end{figure}

\begin{figure}
\centering
\includegraphics[angle=0.0,scale=1.0]{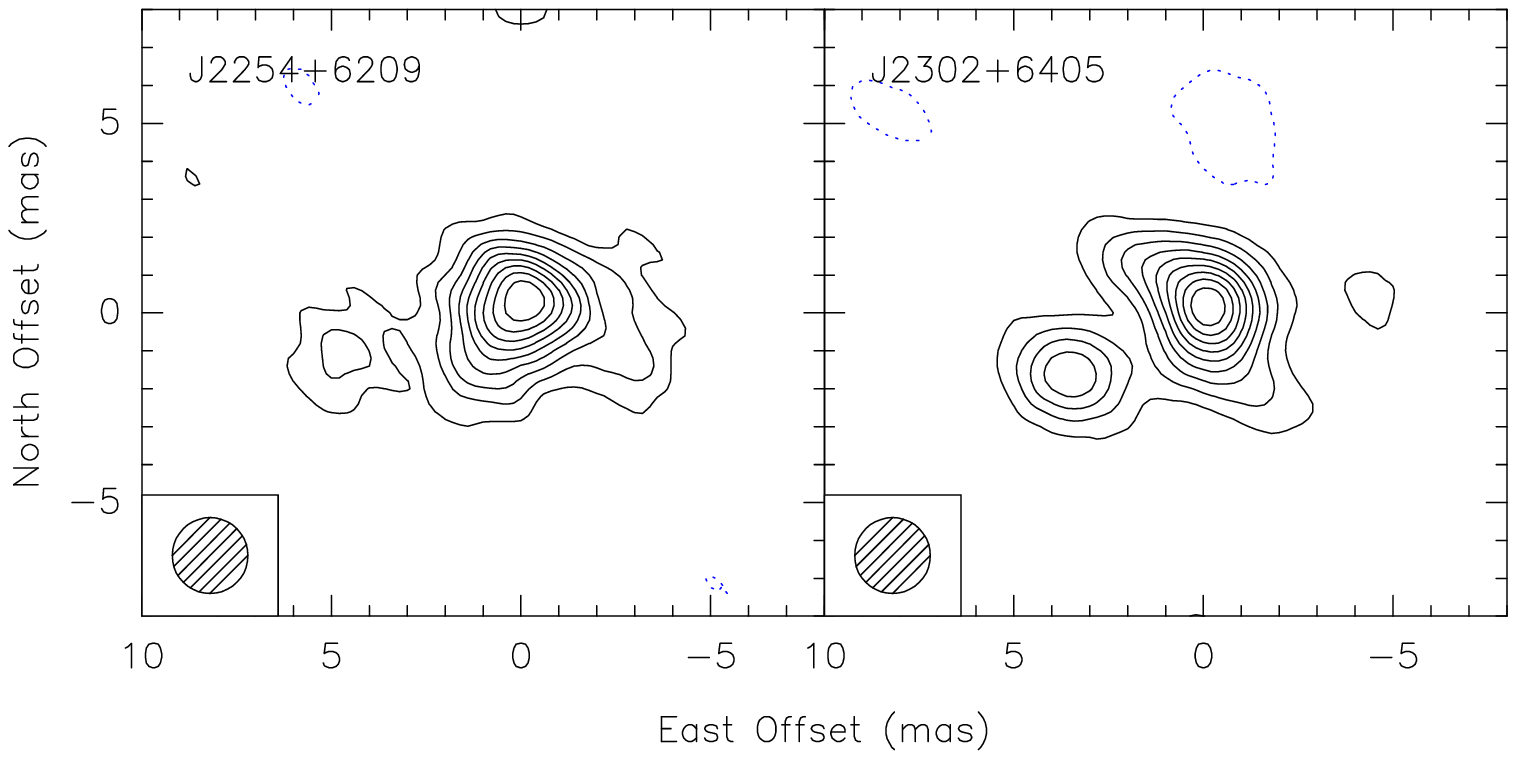}
\caption{Images of the two background continuum sources phase referenced to the Cep~A maser. 
Source names are in the upper left corner and restoring beams are in the lower
left corner of each panel. Both images are from 
the first epoch observations on 2005 Sep. 9.  
Contour levels of both images are at multiples of 10\% of the peak brightness
of 0.03~Jy~beam$^{-1}$ and 0.04~Jy~beam$^{-1}$ 
for the J2254+6209 and the J2302+6405 image, respectively. 
\label{Cep_images_2}}

\end{figure}

Since the brighest maser emission from Cep~A was only $\approx1$~Jy~beam$^{-1}$
(see Fig.~\ref{Cep_images_1}),
the reference-phase solutions were marginal and this resulted in 
poor image quality for the two background sources (see Fig.~\ref{Cep_images_2}), 
limiting the accuracy of position fits.  
Because the maser spots were fairly weak and possibly blended
(see Table~\ref{CepheusA_spot}), we decided against image-plane
fitting.  As for \NGC, we simply used the centroid position of the maser 
reference channel for parallax fitting.  Formal fitting uncertainties 
for the background continuum sources were 
$\sim0.05$~mas for these images.

\begin{deluxetable}{ccrrr}
\tablecolumns{6} \tablewidth{0pc} \tablecaption{Cep~A : Parallax \& Proper Motion Fit}
\tablehead {
  \colhead{Maser V$_{\rm LSR}$} & \colhead{Background} &  \colhead{Parallax} &
  \colhead{$\mu_{x}$} & \colhead{$\mu_{y}$} 
\\
  \colhead{(km~s$^{-1}$)}  &  \colhead{Source} &  \colhead{(mas)} &
  \colhead{(mas~y$^{-1}$)} & \colhead{(mas~y$^{-1}$)} 
            }
\startdata
 $-$4.2 & J2254+6209      &  1.34$\pm$0.10 &  1.7$\pm$2.0 &  $-$3.8$\pm$0.2 \\
 $-$4.2 & J2302+6405      &  1.51$\pm$0.11 &  $-$0.8$\pm$0.8 &   $-$3.6$\pm$0.2 \\
& & & & \\
 $-$4.2 & \mbox{combined} & 1.43$\pm$0.08 &  0.5$\pm$1.1  &  $-$3.7$\pm$0.2 \\
\enddata
\tablecomments {Col.~1 reports the LSR velocity of the reference maser channel; 
Col.~2 indicates the background quasar whose data were used for the parallax fit: "combined" means that both quasars' data were used; 
Col.~3 reports the fitted parallax; Cols.~4~and~5 give the fitted proper motions along the North and East direction, respectively.}
\label{par_fit_CEPA}
\end{deluxetable}

\begin{figure}
\includegraphics[angle=-90,scale=0.7]{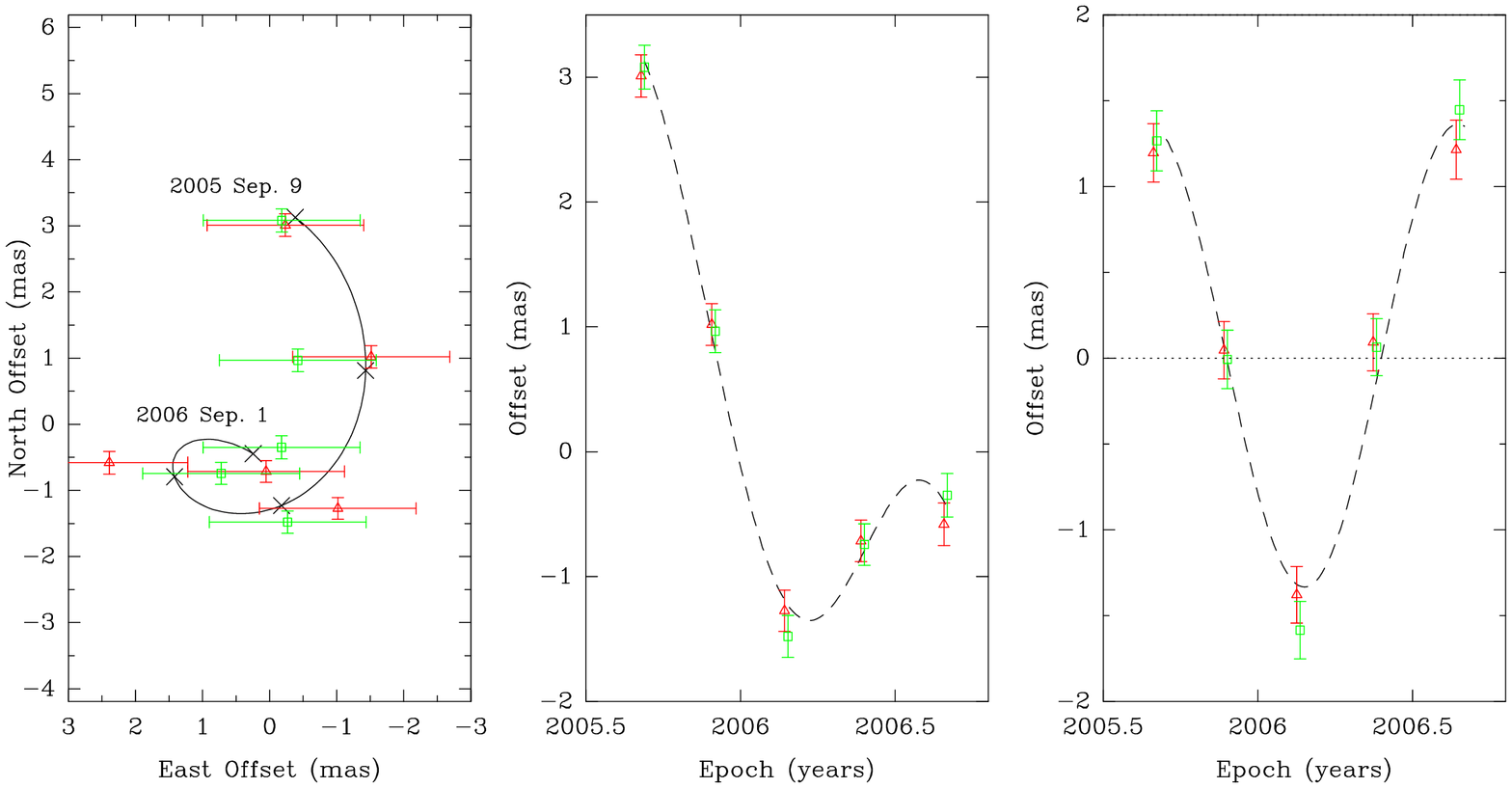}
\caption{Results of the parallax fit for Cep~A.  For each plot, 
colored symbols indicate the position offset of the
reference maser channel centroid; the errorbars have been
scaled to give a reduced $\chi^2$ of unity, as discussed in the text.
{\it Red triangles} and {\it green squares} refer to measurements
relative to the background continuum sources J2254+6209 and J2302+6405, 
respectively.  
{\bf Left Panel:}  Sky-projected motion of the maser.
The {\it crosses} and the {\it continuous line} show the best-fit position 
offsets and the trajectory, respectively.  The observing date of the 
first and last position are indicated. 
{\bf Middle Panel:}  The position offsets of the maser channel along the 
North direction are plotted versus time, with the {\it dashed} line showing 
the best-fit model. The eastward offsets had large ($\ge1$~mas) 
uncertainties and are not shown. 
{\bf Right Panel:} Same as for the middle panel, but with the fitted slope 
(proper motion) subtracted. The {\it dotted} line indicates zero position offset.
\label{CepheusA_parallax} }
\end{figure}

For Cep~A, we perfomed the parallax fit separately for the two 
background quasars and also a combined solution.
The error floors required to give fits with unity reduced $\chi^2$ 
were  1.22~mas and 0.17~mas.  The large eastward error floor for
Cep~A probably is caused by the large east-west extension of the maser
and likely blending problems, and the Cep~A parallax is effectively 
determined only by the north-south data.  Table~\ref{par_fit_CEPA} reports 
results of these fits and Fig.~\ref{CepheusA_parallax} displays the data
and the best fitting models. 
The errorbars for the positions in this figure include the error floors.

\section{Discussion}

\subsection {Galactic Locations and Peculiar Motions}

For both maser source, \NGC\ and Cep~A, our parallax distances are accurate 
to $\approx5$\%.  This is significantly better than the accuracy of 
photometric distances for these sources, which are typically accurate to 
10--20\% \citep{Joh57,Mor86}.  

The parallax distance of \NGC\ of 
$2.65^{+0.12}_{-0.11}$~kpc is  within the range of values 
reported in the literature: from 2.2~kpc \citep{Mor86} to 2.8~kpc
\citep{Cra78}.  
The kinematic distance of \NGC\ is $5.6$~kpc, assuming standard
values for the rotation of the Galaxy ($R_0=8.5$~kpc and $\Theta_0=220$~\kms).
At this distance, \NGC\ would be well beyond the Perseus spiral arm,
possibly in an ``Outer'' (``Cygnus'') arm.  However, the parallax distance 
is about a factor of two smaller than its kinematic distance, placing \NGC\ in the 
Perseus spiral arm.

For Cep~A, the methanol parallax distance of  $0.70^{+0.04}_{-0.04}$~kpc 
is consistent with the most-cited value in the literature of 0.725~kpc 
\citep{Joh57}.   The distance to Cep~A is also smaller than its kinematic
distance of 1.1~kpc and places it in the ``Local'' (or ``Orion'') spur.

Combining the distances, LSR velocities and proper motions of the
masers yields their locations in the Galaxy and their full space motions.  
Since internal motions of 12~GHz methanol masers are fairly small, 
typically $\sim3$~\kms \citep{Mos02}, the maser motions should be close 
to that of their associated young stars.  Given a model for the scale and 
rotation of the Milky Way, we can subtract the effects of Galactic rotation 
and the peculiar motion of the Sun from the space motions of the 
maser sources and estimate the peculiar motions of the maser 
star forming regions.  We adopt the IAU values for the distance to 
the Galactic center ($R_0=8.5$~kpc) and the rotation speed of the 
Galaxy at this distance ($\Theta_0=220$~\kms) and the Hipparcos 
measurements of the Solar Motion \citep{Deh98}.
For these parameters and a flat rotation curve, the peculiar velocity components for \NGC\ are 
$(U_s,V_s,W_s)=(25\pm2,-30\pm3,-10\pm1)$~\kms\ and for Cep~A are 
$(5\pm3,-12\pm3,-5\pm2)$~\kms, where $U_s, V_s \, {\rm and} \, W_s$ are velocity 
components toward the Galactic center, in the direction of Galactic rotation, 
and toward the North Galactic Pole, respectively, at the location of the source.  
The uncertainties for the peculiar motions reflect measurement
errors for parallax, proper motion, and $V_{\rm LSR}$ ($\pm3$~\kms assumed),
but no systematic contribution from uncertainty in the Galactic model or
Solar Motion.
 
\NGC\ has large peculiar velocity components toward the Galctic center and
counter to Galactic rotation.  Cep~A also has a significant peculiar motion
counter to Galactic rotation.  The implications of these peculiar velocities 
for models of Galactic rotation and structure will be discussed in a later 
paper, based on results for a large number of maser sources.

\subsection{12~GHz maser spatial distribution}

Tables~\ref{NGC7538_spot}~and~\ref{CepheusA_spot} give the strengths
and velocities of maser spots detected in \NGC\  and Cep~A, respectively. 
% The absolute position of the reference maser spots (label number 1)
%are given in Table~\ref{obs_sou}.
The accuracy of the relative positions of maser spots within each
source is usually limited by the complex spatial and velocity distribution 
of the maser emission and is typically 0.1 to 0.5~mas.  At the distances of 
the masers and over our time baseline of 1~yr, this positional accuracy 
leads to an uncertainty in relative velocities of 2 to 7~\kms\ for \NGC\  
and 0.4 to 2~\kms\ for Cep~A.  As shown in Tables~\ref{NGC7538_spot} and
\ref{CepheusA_spot}, internal motions are small (mostly $<5$~\kms).

\begin{deluxetable}{rrrcrrcrr}
\tablecolumns{6} \tablewidth{0pc} \tablecaption{Parameters of methanol 12~GHz maser spots detected in \NGC\ }
\tablehead {
 \colhead{Label} & \colhead{V$_{\rm LSR}$} & \colhead{$F_{\rm int}$} &  & 
 \colhead{$\Delta \alpha $ } &  \colhead{$\Delta \delta$ } & & 
\colhead{$V_{\rm x}$} &  \colhead{$V_{\rm y}$} \\
   & \colhead{(km~s$^{-1}$)}  &  \colhead{(Jy)} & &
 \colhead{(mas)} & \colhead{(mas)} & &
  \colhead{ (km s$^{-1}$) } & \colhead{ (km s$^{-1}$) }
            }
\startdata
  1 & $-$55.8 &  18.8 & &      0   &     0   & &     0   &      0   \\
  2 & $-$56.5 &   9.3 & & $-$5.9$\pm$0.4  & 1.7$\pm$0.1  & & $-$3$\pm$7 
& 1$\pm$2   \\
  3 & $-$55.8 &   3.8 & & 4.7$\pm$0.4  & $-$2.2$\pm$0.1  & &     1$\pm$7 
&     5$\pm$2    \\
  4 & $-$55.8 &   2.7 & &     8.2$\pm$0.4  &     -3.9$\pm$0.1  & &     8$\pm$7 
&   $-$4$\pm$2    \\
  5 & $-$61.1 &   0.8 & &   $-$92.1$\pm$0.4  &   $-$225.9$\pm$0.1  & &  &  \\
  6 & $-$61.1 &   0.7 & &    $-$72.0$\pm$0.4  &   $-$215.8$\pm$0.1  & &    $-$3$\pm$7 
&    1$\pm$3   \\
  7 & $-$61.1 &   0.6 & &   $-$88.3$\pm$0.4  &   $-$223.5$\pm$0.1  & &  &  \\
  8 & $-$56.5 &   0.6 & &  $-$35.3$\pm$0.4  &     11.5$\pm$0.2  & &     3$\pm$8 
&   0$\pm$3   \\
 9 & $-$56.5 &   0.4 & &    $-$38.9$\pm$0.5  &      9.0$\pm$0.2  & &     1$\pm$8 
&     0$\pm$3   \\
  10 & $-$61.1 &   0.3 & &   $-$75.2$\pm$0.4  &   $-$219.1$\pm$0.1  & &    $-$1$\pm$7
&     6$\pm$3  \\
 11 & $-$58.1 &   0.3 & &    118.6$\pm$0.4 &   $-$178.9$\pm$0.2  & &     0$\pm$8 
&     4$\pm$3    \\
 12 & $-$57.3 &   0.1 & &    $-$71.8$\pm$0.4  &     20.1$\pm$0.2  & &  &  \\
 13 & $-$58.1 &   0.1 & &    124.7$\pm$0.5  &   $-$176.7$\pm$0.3  & &  &  \\
 14 & $-$57.3 &   0.1 & &    $-$77.0$\pm$0.4  &     21.6$\pm$0.2  & &  &  \\
 15 & $-$58.1 &   0.1 & &    121.4$\pm$0.5  &   $-$169.4$\pm$0.2  & &  &  \\
 16 & $-$61.9 & 0.02 & &   $-$282.3$\pm$0.4  &    124.0$\pm$0.2  & &  &   \\
\enddata
\tablecomments {For each identified spot, Col.~1 reports the label number, increasing with decreasing spot intensity; Cols.~2 and ~3  the LSR velocity  and  the integrated flux density; Cols.~4 and ~5  the (eastward and northward) positional offsets evaluated with respect to the spot with label number 1;
Cols.~6 and ~7 the projected components along the East and North direction of the proper motion relative to the spot with label number 1.
The derived absolute position of the maser reference spot (label number 1) is: R.A.(J2000) = 23$^{\rm h}$ 13$^{\rm m}$  45$\fs$3622, Dec.(J2000) = 61$\degr$ 28$^{\prime}$ 10$\farcs$507.}
\label{NGC7538_spot}
\end{deluxetable}

\begin{deluxetable}{rrrcrrcrr}
\tablecolumns{6} \tablewidth{0pc} \tablecaption{Parameters of methanol 12~GHz maser spots detected in Cep~A }
\tablehead {
 \colhead{Label} & \colhead{V$_{\rm LSR}$} & \colhead{$F_{\rm int}$} &  & 
 \colhead{$\Delta \alpha $ } &  \colhead{$\Delta \delta$ } & & 
\colhead{$V_{\rm x}$} &  \colhead{$V_{\rm y}$} \\
   & \colhead{(km~s$^{-1}$)}  &  \colhead{(Jy)} & &
 \colhead{(mas)} & \colhead{(mas)} & &
  \colhead{ (km s$^{-1}$) } & \colhead{ (km s$^{-1}$) } 
            }
\startdata
 1 &  $-$4.2 &   1.8 & &      0  &      0  & &      0   &      0     \\
  2 &  $-$4.2 &   1.1 & &    4.8$\pm$0.4  &     $-$0.1$\pm$0.2  & &  &  \\
  3 &  $-$1.9 &   1.1 & &  $-$1348.3$\pm$0.5  &    150.4$\pm$0.3  & &    $-$6$\pm$3 
&   2$\pm$1  \\
  4 &  $-$4.2 &   0.9 & &   $-$4.8$\pm$0.4  &     $-$0.2$\pm$0.2  & &    $-$1$\pm$2 
&     1$\pm$1   \\
  5 &  $-$4.2 &   0.5 & & $-$1168.0$\pm$0.5  &  $-$147.0$\pm$0.4  & &  &  \\
  6 &  $-$1.9 &   0.3 & &  $-$1345.5$\pm$0.5  &    151.4$\pm$0.2  & &  &  \\
  7 &  $-$1.9 &   0.3 & &  $-$1355.1$\pm$0.6  &    150.6$\pm$0.4  & &  &   \\
  8 &  $-$4.2 &   0.2 & &    19.4$\pm$0.5 &      1.8$\pm$0.2  & &  &  \\
\enddata
\tablecomments {For each identified spot, Col.~1 reports the label number, increasing with decreasing spot intensity; Cols.~2 and ~3  the LSR velocity  and  the integrated flux density; Cols.~4 and ~5  the (eastward and northward) positional offsets evaluated with respect to the spot with label number 1;
Cols.~6 and ~7  the projected components along the East and North direction of the proper motion relative to the spot with label number 1.
The derived absolute position of the maser reference spot (label number 1) is: R.A.(J2000) = 22$^{\rm h}$ 56$^{\rm m}$  18$\fs$0970, Dec.(J2000) = 62$\degr$ 01$^{\prime}$ 49$\farcs$399.}
\label{CepheusA_spot}
\end{deluxetable}

\subsubsection  {Cep~A}

\begin{figure}
\centering
\includegraphics[angle=-90,scale=0.9]{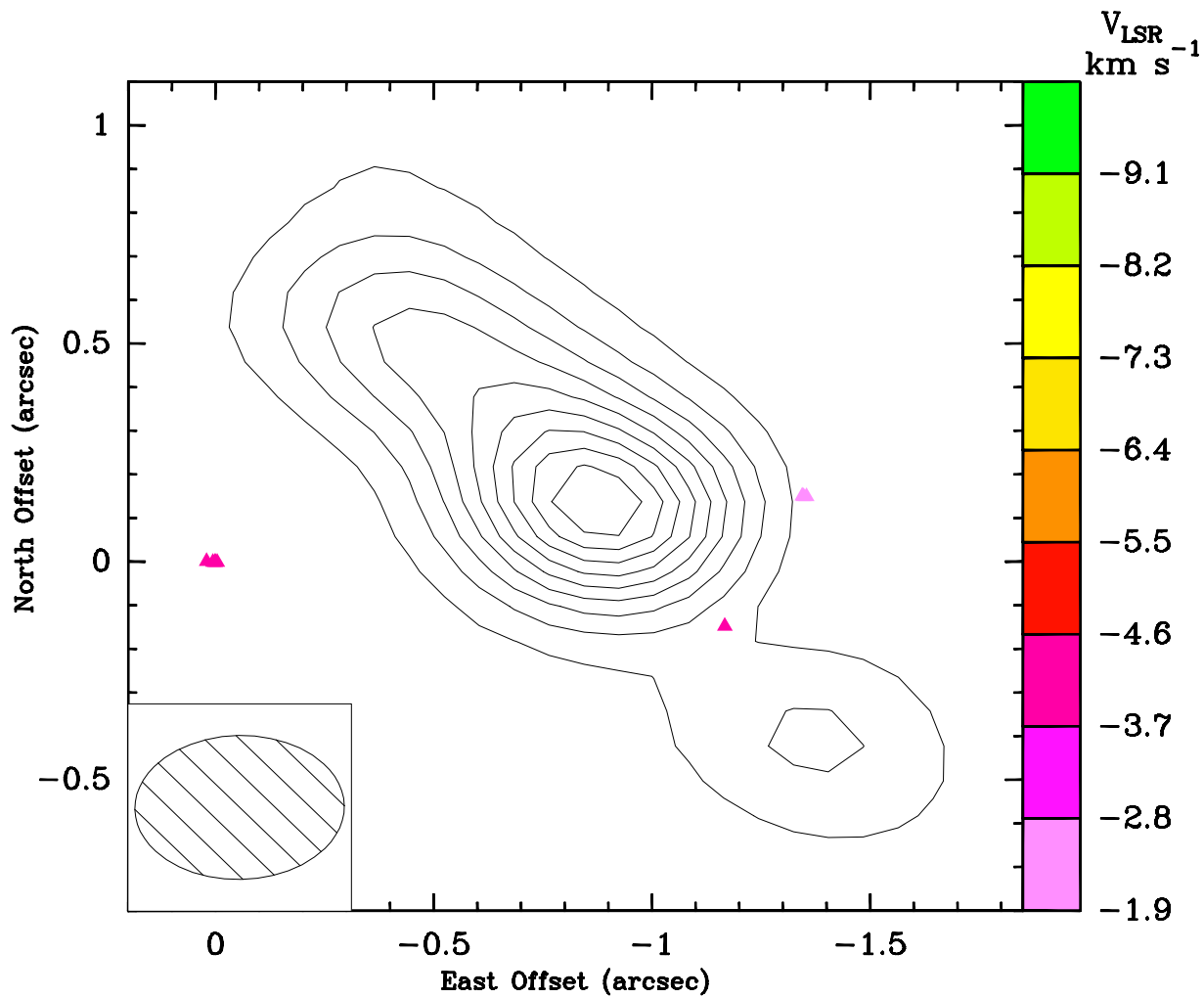}
\caption{\small Cep~A methanol 12~GHz maser distribution 
({\it filled triangles}) plotted on an 8.4~GHz continuum VLA A-configuration 
image ({\it contours}) of the Cep~A HW2 jet. 
Plotted levels of the continuum emission are  
at multiples of 10\% of the peak brightness  
of \ 5~mJy~beam$^{-1}$. 
Positions are relative to the most intense 12~GHz maser spot 
(label number 1 in Table~\ref{CepheusA_spot}). 
Different {\it colors} are used to indicate the maser LSR velocities, 
according to the color scale on the right-hand side of the plot. 
The color scale was chosen with the source rest velocity set to green
and the maximum maser velocity set to purple.
%Spot (relative) proper motions are affected by large uncertainties 
%(see Table~\ref{CepheusA_spot}) and are not shown. 
The insert at the bottom left corner shows the FWHM beam size 
of the VLA image. Maser spots in this source are weak and their 
velocity-averaged emission maps show little structure. 
%{\bf Insert Panel:} The small panel in the upper right corner of the main panel
%shows the velocity-averaged emission of the group of more intense spots 
%(denoted with label numbers 1, 2~and~4 in Table~\ref{CepheusA_spot}) 
%at the first observing epoch (2005 Sep. 9). 
%Plotted levels are at multiples of 10\% 
%of the  (velocity-averaged) spot peak intensity of 0.5~Jy~beam$^{-1}$.
\label{CepheusA_distri}
}
\end{figure}

Fig.~\ref{CepheusA_distri} shows the spatial distribution of the 
12~GHz methanol masers detected toward Cep~A overlaid on a 
map of the Cep~A HW2 jet \citep{Tor96}.
The map of the Cep~A HW2 jet was made by analyzing
NRAO Archive VLA data from program AC534 observed in 1999.
All spots concentrate in three
clusters separated by 1$\farcs$35 and 0$\farcs$15 along the
East and North direction, respectively.
Two of the 12~GHz methanol maser spots appear to match
in separation with those mapped by \citet{Min00} from
VLBA observations in January 1999.  However, one of
Minier's spots does not appear in our map and, conversely,
one of ours does not appear in his map.  This indicates
a time scale of order a decade for significant flux density
variations.   The detected maser clusters fall on both sides of
the Cep~A HW2 YSO, which is the most massive member of a
group of protostellar objects inside a region of radius
$\approx1\arcsec $ \citep{Com07}.  Given the complexity of
this star-forming region, it is unclear whether the
12~GHz methanol masers are excited by a single or
multiple protostellar objects.

\subsubsection{\NGC}

\begin{figure}
\includegraphics[angle=-90,scale=0.9]{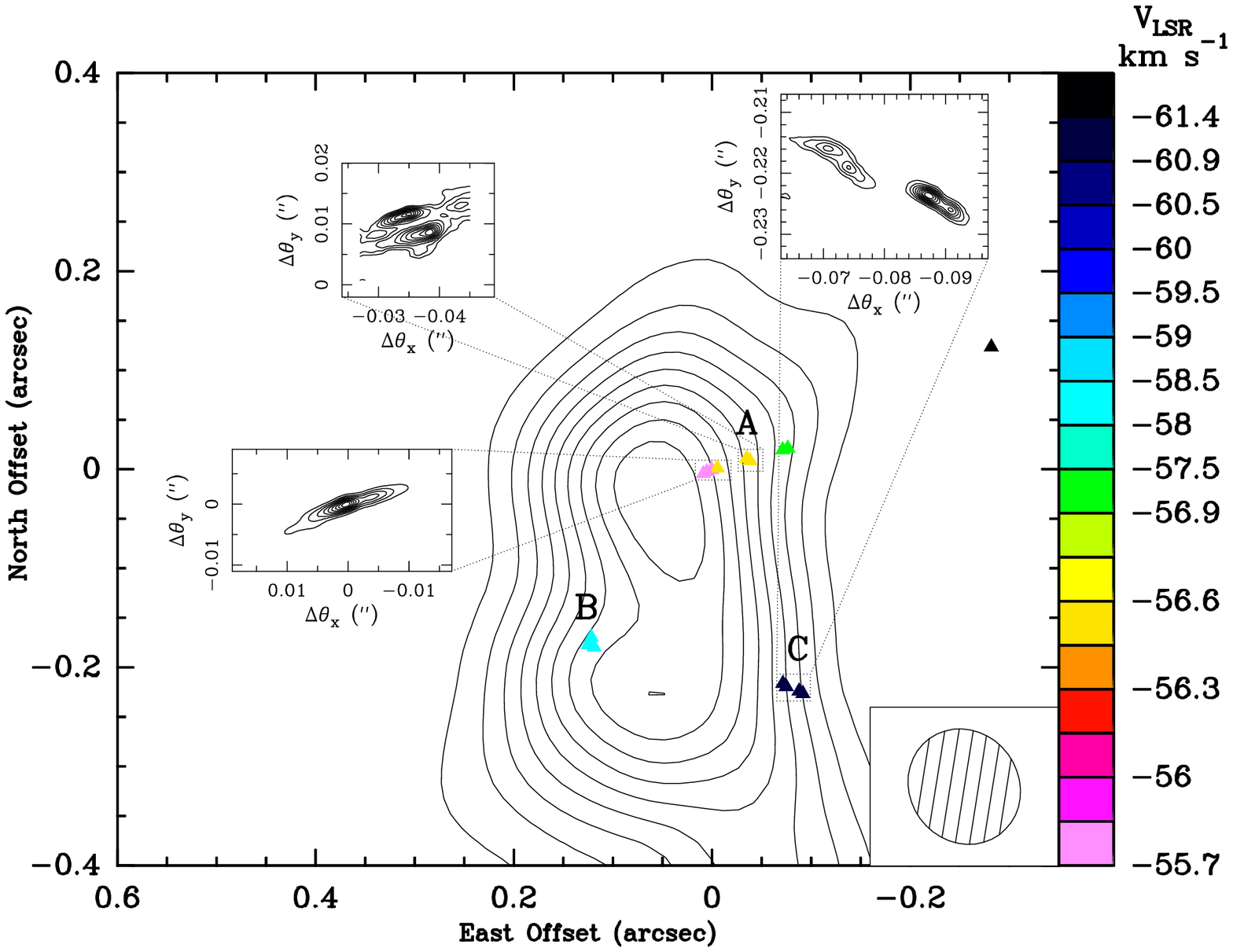}
\caption{\small \NGC\ 12~GHz methanol maser distribution  
({\it filled triangles}) plotted on a 15~GHz continuum VLA A-configuration
image ({\it contours}) of \NGC\ IRS1. Plotted levels of the continuum
emission are at multiples of 10\% of the peak brightness  
of \ 23~mJy~beam$^{-1}$.
Positions are relative to the most intense 12~GHz maser spot, 
labeled number 1 in Table~\ref{NGC7538_spot}. 
Different {\it colors} are used to indicate the maser LSR
velocities, according to the color scale on the right-hand side
of the plot.  The 6.7 and 12~GHz maser clumps previously 
detected by \citet[Fig.~1]{Min00}
are labeled with capital letters A, B and C. 
%Spot (relative) proper motions are affected by large uncertainties (see Table~\ref{NGC7538_spot}) and are not shown. 
The insert on the bottom right corner shows the FWHM beam size 
of the VLA image. 
{\bf Small Panels:} The three small panels around the continuum image
show the velocity-averaged emission of different groups of intense maser spots 
at the first observing epoch (2005 Sep. 9). 
Plotted levels are at multiples of 10\% of the  
(velocity-averaged) spot peak intensity, corresponding to
1.6~Jy~beam$^{-1}$, 0.13~Jy~beam$^{-1}$ and 0.2~Jy~beam$^{-1}$
for spots shown in the lower left, upper left and upper right panel, respectively.
 \label{NGC7538_distri}}
\end{figure}

Fig.~\ref{NGC7538_distri} shows the spatial distribution of
the 12~GHz methanol masers detected toward \NGC\ overlaid on
a 15~GHz VLA A-configuration map of the ultra-compact HII region
IRS~1 \citep{Gau95}.   The continuum map was produced by
analyzing NRAO Archive VLA data (program AF0413) from
observations  in 2004.  The color code for the maser emission
in Fig.~\ref{NGC7538_distri} is centered (green) on the a
systemic LSR velocity of $-57.0$~\kms, as indicated by high
spectral resolution observations of the molecular emission
from this core \citep{Kam88}.

Toward \NGC\ we detected 16 maser spots, of which 9 persisted
throughout our observations.  Comparing to previous methanol
6.7 and 12~GHz methanol observations of \citet[Fig.~1]{Min00}, 
we find his maser clusters A, B and C.
We do not detect clusters D or E, which were previously
identified only at 6.7 GHz.  However, we do find a 12 GHz
methanol maser spot at a (North, East) offset of ($-$0$\farcs$23,0$\farcs$13)
that does not appear in Minier's map.

We confirm the possible velocity gradient seen in cluster A
of \citet{Min00}, indicating that this velocity/position
structure is stable over a timespan of at least 7 years.
This structure has been interpreted as either an edge-on
rotating disk \citep{Min00,Pes04, Pes04b} or a collimated
outflow \citep{DeB05}.  The expected proper motions for these
two models would be quite different and they may offer a
method to discriminate between these models.   However, since
the spread in radial velocities across cluster A is only about
$3$~\kms, one would like proper motions with accuracies
better than $1$~\kms ($<0.1$~mas~y$^{-1}$).  Future observations of
the 12 GHz methanol masers with the VLBA should yield proper
motions with such accuracies.

\acknowledgments
\noindent
Andreas Brunthaler was supported by the DFG Priority Programme 1177. \\
Ye Xu was supported by Chinese NSF through grants NSF 10673024, NSF 10733030, NSF 10703010 and NSF 10621303.

%\bibliographystyle{aa}
%\bibliography{biblio}

{\it Facilities:} \facility{VLBA}.

%\appendix

\end{document}